\documentclass[final,leqno,onefignum,onetabnum]{siamltex1213}

\usepackage{amssymb,amsmath}

\usepackage[capitalise]{cleveref}

\def\cbot{c_{\text{bot}}}
\def\cmin{c_{\text{min}}}

\def\lam{\lambda}
\def\R{\mathbb{R}}
\def\Z{\mathbb{Z}}
\def\calP{\mathcal{P}}
\def\eg{{e.\,g.}}
\def\ie{{i.\,e.}}
\def\st{\text{s.\,t.}}
\def\barc{\bar{c}}
\def\eprime{f}
\def\NP{N\!P}

\def\flowproblem{FP} 
\def\designproblem{DP} 
\def\combinedproblem{CP} 
\def\flowlp{[\textup{LP}_{\textup{flow}}]}
\def\uniformcostlp{[\textup{LP}_{\textup{design}}]}

\title{Protection of flows under targeted attacks}

\author{%
Jannik Matuschke\thanks{Institut f{\"u}r Mathematik, Technische Universit\"{a}t Berlin (\email{matuschke@math.tu-berlin.de}).} \and 
S. Thomas McCormick\thanks{Sauder School of Business, University of British Columbia (\email{tom.mccormick@sauder.ubc.ca}).} \and 
Gianpaolo Oriolo\thanks{Dipartimento di Ingegneria Civile e Ingegneria Informatica, Universit\`a di Roma ``Tor Vergata'' (\email{oriolo@disp.uniroma2.it}).} \and \newline
Britta Peis\thanks{Fakult\"at f\"ur Wirtschaftswissenschaften, RWTH Aachen (\email{britta.peis@oms.rwth-aachen.de}).} \and 
Martin Skutella\thanks{Institut f{\"u}r Mathematik, Technische Universit\"{a}t Berlin} (\email{martin.skutella@tu-berlin.de}).}

\begin{document}
\maketitle

\begin{abstract}
Due to the importance of robustness in many real-world optimization problems, the field of robust optimization has gained a lot of attention over the past decade. We concentrate on maximum flow problems and introduce a novel robust optimization model which, compared to known models from the literature, features several advantageous properties: (i) We consider a general class of path-based flow problems which can be used to model a large variety of network routing problems~(and other packing problems). (ii) We aim at solutions that are robust against targeted attacks by a potent adversary who may attack any flow path of his choice on any edge of the network. (iii)~In contrast to previous robust maximum flow models, for which no efficient algorithms are known, optimal robust flows for the most important basic variants of our model can be found in polynomial~time.

We also consider generalizations where the flow player can spend a budget to protect the network against the interdictor. Here, we show that the problem can be solved efficiently when the interdiction costs are determined by the flow player from scratch. However, the problem becomes hard to approximate when the flow player has to improve an initial protection infrastructure. 
\end{abstract}

\section{Introduction}\label{sec:intro}

\emph{Network flow problems} form one of the most important classes of optimization problems with numerous real-world applications, \eg, in production systems, logistics, and communication networks. The increasing dependence of our society on constant availability of such network services motivates the study of new flow models that are \emph{robust} against unforeseen interferences, link failures, and targeted attacks by external forces.

The theory of \emph{robust optimization} offers various techniques to handle the issue of planning in face of uncertainties and unreliability; see, \eg,~\cite{Ben-TalElGhaouiNemirovski2009,BertsimasBrownCaramanis2011} for surveys.
A general idea is to model uncertainty by a set of possible scenarios~$\Omega$ that is specified along with the instance of the optimization problem under consideration, where each scenario represents a possible outcome involving failures in the infrastructure, intentional sabotage, or similar complications. With respect to a worst-case analysis, the \emph{robust} objective value of a feasible solution $x$ is the worst
objective value of $x$ among all possible scenarios $z \in \Omega$.

Robust optimization can thus be interpreted as a two-player game: the first player (``the decision-maker'') chooses a solution $x$ from a set $X$ of feasible solutions to the
underlying ``nominal'' optimization problem. Afterwards, the second player (``the adversary'' or ``interdictor'') selects a scenario $z$ from the predefined scenario set $\Omega$.
While the first player aims at maximizing the resulting objective value $\mbox{val}(x,z)$, the adversary selects $z\in \Omega$ in order to reduce $\mbox{val}(x,z)$ as far as possible.
The robust optimization problem  therefore asks for an optimal $x\in X$ solving
$$\max_{x\in X} \min_{z\in \Omega} \mbox{val}(x,z).$$

In this paper, we present a new robust optimization model for network flows.
In existing models the interdictor acts on a subset of the arcs of the network and the interdiction of an arc affects all flow on that arc equally.
By contrast, our model allows the interdictor to specify the amount of flow removed from each flow path individually (we therefore deal with flows on {\em paths}). In this context, it might be helpful to think of the interdictor as a thief who steals particular flow units of his choice: As an illustrative example, consider a train network in which each flow path represents a train and train robbers try to remove as much cargo as possible from the trains, attacking each train at the most vulnerable point it traverses.
Besides providing this new perspective of robust flow optimization, the new model has the advantage that optimal robust flows can be computed in polynomial time.
We also consider further variants of the problem, in which the flow player can adjust the protection for the flow he sends through the network on each arc, subject to a budget. 
While we focus our discussion throughout this paper mainly on the classic maximum flow problem, we also point out that our results directly extend to robust optimization versions of a general class of packing problems that extends beyond network flows.

\subsubsection*{Contribution and structure of the paper}
In the remainder of this section, we present our new robust flow model and compare it to existing robust flow models. We also discuss literature on the closely related field of network interdiction.

In \cref{sec:flow}, we study the basic version of our new robust flow model, in which interdiction costs for the arcs of the network are given and the flow player determines a path flow with the goal of maximizing the surviving flow value after interdiction. 
We show that the optimal strategy for the flow player can be found by solving a parametric LP, where the parameter corresponds to the cost of the most expensive arc affected by the interdictor. We also show that in general, optimal solutions to our problem are not integral, and that any combinatorial algorithm for our problem can also be used to solve a feasibility version of the multicommodity flow problem combinatorially. Finally, we point out that our results still hold when the flow player's options are limited by a budget, and further extend to a very general class of packing problems, including multicommodity flows, abstract flows, and $b$-matchings. 

In \cref{sec:design}, we study a design variant of the problem, where the flow player has to buy the protection of the flow he sends through the network subject to a limited budget. For each arc, the cost of protection is proportional to the chosen interdiction cost on that arc and the flow that needs to be protected. We show that this seemingly hard non-linear optimization problem can be solved by exploiting insights on the structure of an optimal solution.

In \cref{sec:combined}, we discuss a generalization of the problem from the preceding section, in which an initial (free) protection of the flow on each arc is given but the  interdiction costs can be further increased by the flow player subject to his budget. We show, that in contrast to the problems discussed earlier, this problem is not only $N\!P$-hard but does not even allow for approximation algorithms.

\subsection{The new model}
We are given a directed graph $D=(V,A)$ with source $s\in V$ and sink $t\in V$, and arc capacities $u \in \R_+^A$.
We consider flows on paths, as in \cite[Section~4]{BertsimasNasrabadiStiller2013}, and so 
let $\calP$ denote the collection of all $s$-$t$-paths in $D$.
The strategy choices of the decision maker (whom we call the \emph{flow player}) are given by the set
$$X:=\Big\{x\in \R^{\mathcal{P}}_+\mid \sum_{P\in \calP: e\in P} x_P \le u_e \ \forall e\in A\Big\}$$
of all feasible $s$-$t$-flows in the capacitated network $(D,u)$, \ie, the flow player specifies the amount of flow along each $s$-$t$-path subject to the arc capacities.

Classic robust flow models (which are discussed further below) are built on the assumption that the interdictor attacks arcs of the network subject to a budget, equally affecting all flow paths traversing the interdicted arcs. In contrast, in our model we instead think of the interdictor as a thief who might directly attack and steal flow on each individual path rather than manipulating an arc $e$ as a whole. 
Each arc $e\in A$ is equipped with an \emph{interdiction cost} $c_e\ge 0$,  specifying the cost of stealing one unit of flow on that arc.
The interdictor can, after the flow player has chosen a flow $x \in X$,
use a given budget $B_I$ in order to steal flow on some of the paths.
Therefore the interdictor chooses a scenario/strategy 
$$z\in \Omega:=\Big\{z\in \R_+^{A\times \calP} \mid \sum_{e\in A} c_e \sum_{P\in \calP:e\in P} z_{e,P} \le B_I\Big\}.$$
The remaining flow after applying the interdiction strategy $z$ to flow $x$ is defined by $$\bar{x}_P := (x_P - \sum_{e \in P} z_{e, P})^+$$ for each $P\in \mathcal{P}$. The goal of the flow player is to maximize $\mbox{val}(x,z):=\sum_{P \in \calP} \bar{x}_P$, anticipating the reply of the interdictor who wants to minimize the same quantity, that is, steal as much flow as possible.

Note that an interdictor's attack on a particular path~$P\in\calP$ should always happen on a cheapest arc~$e\in P$. Therefore, after the flow $x \in X$ has been chosen, an optimal strategy for the interdictor is the following greedy approach: Sort the paths $P\in\calP$ in order of non-decreasing bottleneck cost~$\barc_P:=\min_{e\in P}c_e$ and steal flow along the paths in this order until the budget~$B_I$ has been used up. 

This tractability of the interdictor's optimal strategy is a desirable property of our model as it allows  computation of the robust value of any given flow. Note that, in contrast, for the models in~\cite{BertsimasNasrabadiStiller2013,McCormickOrioloPeis2014} discussed below, the interdictor's optimal answer to a given flow is $N\!P$-hard to compute in general (in both cases, the interdictor's problem is equivalent to the budgeted maximum coverage problem~\cite{BertsimasNasrabadiStiller2013}).

Also, with the exception of the basic model in~\cite{AnejaChandrasekaranNair2001}, no efficient algorithm or constant factor approximation is known for the flow player's problem in the robust flow models discussed below despite intense research. We will show that for our model, both the maximum flow version as well as a network design version (in which the flow player adjusts the protection of the links in the network) can be solved efficiently. Furthermore, our model and these positive algorithmic results naturally extend to a very general class of packing linear programs, including, \eg, multicommodity flows, abstract flows, and $b$-matchings, and allows the easy integration of additional budget constraints.

\subsection{Related work} In the following, we discuss existing robust flow models and the related concepts of network interdiction and fortification games.

\subsubsection*{Robust flows} Robust flows subject to cost uncertainties were  studied by Bertsimas and Sim~\cite{bertsimas2003robust}.  Aneja et al.~\cite{AnejaChandrasekaranNair2001} started the study of robust maximum (path) flows in presence of an interdictor (capacity uncertainty), who in their model could remove a single arc from the network. The goal of the flow player, as in all subsequent papers, is to maximize the value of the surviving flow. Aneja et al.~showed that the problem can be solved in polynomial time using a parametric LP. However, as soon as the interdictor is allowed to remove two arcs, the corresponding dual separation problem becomes $N\!P$-hard as was shown by Du and Chandrasekaran \cite{DuChandrasekaran2007}. On the positive side, Bertsimas et al.~\cite{BertsimasNasrabadiStiller2013} building upon a generalization of the parametric LP used in~\cite{AnejaChandrasekaranNair2001}, gave an LP-based approximation, in terms of the amount of flow removed by the interdictor in an optimal solution, for the case where the interdictor can remove any given number of arcs $B_I$. More recently, Bertsimas et al.~\cite{bertsimas2013power} showed that the same flow also yields a \mbox{$1 + (B_I/2)^2/(B_I+1)$}-approximation.
 Formally, the model considered in \cite{AnejaChandrasekaranNair2001,BertsimasNasrabadiStiller2013,DuChandrasekaran2007} is defined for the uncertainty set~\mbox{$\Omega = \{z\in \{0,1\}^A\mid 1^T z\le B_I\}$} and the flow player aims at solving $\max_{x\in X} \min_{z\in \Omega} \mbox{val}(x,z)$, where $\mbox{val}(x,z) = \sum_{P\in \mathcal{P}} (1-\max_{e\in P} z_e)x_P$ is the amount of flow $x\in X$ that survives after the interdictor selects the scenario $z\in \Omega$. 
A fractional version of this uncertainty set was proposed in~\cite{Burchetal2002}: Each arc has a cost $c_e$ for destroying the entire capacity on $e$ and the interdictor is allowed to fractionally attack the capacities on the arcs, \ie, $\Omega=\{z\in [0,1]^A\mid c^T z\le B_I\}$.
For this setting the LP-based approximation algorithm from~\cite{BertsimasNasrabadiStiller2013} can be combinatorialized using the discrete Newton method~\cite{McCormickOrioloPeis2014}. 

\subsubsection*{Network interdiction and fortification games}
Robust flows are closely related to \emph{network interdiction}, which takes the opposite view of letting the interdictor move first (\ie, destroy a part of the network) and then letting the flow player decide to send the flow within the remaining network.
 Network interdiction has many applications such as, \eg, drug interdiction, protection of networks against terrorism, or hospital infection control and has been studied extensively~\mbox{\cite{altner2010maximum,cormican1998stochastic,phillips1993network,royset2007solving,washburn1995two,wollmer1964removing,wood1993deterministic}}.  In the basic network interdiction problem (NI), an interdictor faces a capacitated network $((V, A), u)$, with two distinguished nodes $s$ and $t$, and aims at choosing a subset $F\subseteq A$ of arcs, whose removal cost $\sum_{e\in F} c_e$ does not exceed his budget $B_I$, so as to minimize the value of the maximum $s$-$t$-flow that can be routed in $((V, A\setminus F), u|_{A\setminus F})$.  In 1993, Phillips~\cite{phillips1993network} proved that the problem is weakly $N\!P$-hard on planar graphs, and gave an FPTAS for planar graphs.  At the same time Wood~\cite{wood1993deterministic} proved that NI is strongly $N\!P$-hard on general graphs, even when $c \equiv 1$ and fractional interdiction of an arc is allowed, \ie, by paying $\alpha_e$, for some $0\leq \alpha \leq 1$, the interdictor can delete $\alpha_eu_e$ units of capacity from arc~$e$. Recently, Baffier and Suppakitpaisarn~\cite{baffier2014approximation} and Bertsimas et al.~\cite{bertsimas2013power} independently derived a $B_I+1$-approximation for the case $c \equiv 1$. For general interdiction costs, Burch et al.~\cite{Burchetal2002} provided a bicriteria pseudo-approximation algorithm when fractional interdiction is allowed.

In later sections, we discuss models in which the flow player can adjust the interdiction cost for flow in the network. The corresponding concept in the context of network interdiction are so-called fortification games, in which the flow player can protect network elements from the interdictor before the latter takes his action; see~\cite{SmithLim2008} for an overview.

\subsubsection*{Weighted abstract flows}
At various places in the analysis of our models we will encounter a weighted version of the maximum flow problem, where each path $P \in \calP$ is associated with a weight $r_P$ specifying the reward per unit of flow send along~$P$:~\mbox{$\max \{\sum_{P \in \calP} r_P x_P \mid x \in X\}$}. Unfortunately it has been known since Pr\"omel~\cite{Promel1982} that it is $N\!P$-hard to find an optimal flow for general weights $r \in \R^{\calP}_+$.  However, Hoffman~\cite{hoffman1974generalization} showed that when the rewards satisfy the following supermodularity property, such problems have integral optimal solutions\footnote{Hoffman's model is more general than this, but it is this supermodularity that will concern us here.}: Whenever $P, Q \in \calP$ share an arc $e$, then there exist paths $P', Q' \in \calP$ such that $P'$ is contained in the first part of $P$ up to $e$ and the last part of $Q$ after $e$, and $Q'$ is the same but with $Q$ first and $P$ second, and 
\begin{equation}
  r_{P'} + r_{Q'} \geq r_P + r_Q.  \label{eq:submod}
\end{equation}
Martens and McCormick \cite{MartensMcCormick2008} call Hoffman's model {\em Weighted Abstract Flow (WAF)}, and developed an oracle-polynomial algorithm for WAF.

\section{Solving the robust flow model}\label{sec:flow}
As mentioned above, we are given a directed graph \mbox{$D=(V,A)$} with arc capacities $u \in \R_+^A$ and costs $c \in \R_+^A$, and a budget~$B_I$. Let $s, t \in V$ be a source and a sink and let $\calP$ denote the set of all $s$-$t$-paths in $D$ (at the end of this section we will also discuss other set systems for which our results hold).
The task of the flow player is to find a feasible path-flow $x \in \R_+^\calP$ with $\sum_{P\in\calP:e\in P}x_P\leq u_e$ for all~$e\in A$. Then the interdictor can use his budget~$B_I$ to steal flow along paths~$P\in\calP$ at cost $\barc_P = \min_{e \in P} c_e$ per flow unit. The goal of the flow player is to maximize his \emph{profit}, \ie, the amount of flow that remains, while the interdictor tries to minimize this quantity, \ie, steal as much as possible. We denote the resulting bilevel optimization problem by {\flowproblem}.

Again recall that the interdictor's optimal strategy is to be greedy, \ie, sort the paths $P\in\calP$ by non-decreasing $\barc_P$ and steal flow along the paths in this order until the budget~$B_I$ has been used up.
The flow player's optimal strategy is less obvious. To compute it, we define the following LP. For a fixed arc $\eprime\in A$, let~$B':=B_I/c_{\eprime}$, $c'_e:=\min\{c_e/c_{\eprime},1\}$, for $e\in A$, and $\barc'_P:=\min_{e\in P}c'_e$ for $P\in\calP$. Consider the linear program
\begin{align*}
\flowlp \qquad \max\quad&\sum_{P\in\calP}\barc'_P x_P-B'\\
\st\quad&\sum_{P\in\calP:e\in P}x_P\leq u_e&&\text{for all~$e\in A$,}\\
&x_P\geq0 && \text{for all~$P\in\calP$.}
\end{align*}

The intuition behind $\flowlp$ is as follows. Assume that $f$ is the most expensive arc that the interdictor touches in his greedy strategy when attacking flow $x$. 
Then for any flow path $P$ that still carries flow after the interdiction, we have that $\barc'_P = 1$, \ie, the total interdiction cost of the surviving flow with respect to $\barc'$ is equal to its value. Furthermore, the interdiction cost of all flow stolen by the interdictor is scaled by the same value as the interdictors budget, \ie, the interdictor removes flow equivalent to a total cost of $B'$ from $x$, motivating the term $-B'$ in the objective function. $\flowlp$ thus asks for a flow maximizing the value of the surviving flow assuming that $f$ is the most expensive arc touched by the interdictor in a pair of optimal strategies. The proof of the following theorem formalizes this intuition and shows that it indeed yields the optimal strategy of the flow player.

\smallskip
\begin{theorem}\label{theorem:optimality}
Solving $\flowlp$ for every fixed arc $\eprime\in A$ and taking the solution with the largest objective function value is an optimal strategy for the flow player.
\end{theorem}
\smallskip

\begin{proof}
Consider an arbitrary flow $x \in X$ and arc $\eprime\in A$. We first show that $\sum_{P \in \calP} \barc'_P x_P - B'$ is a lower bound on the flow player's profit in the original problem when choosing flow $x$ as his strategy. Consider an optimal response $z$ of the interdictor and let $\Delta_P = \sum_{e \in P} z_{e, P}$ denote the amount of flow stolen from path $P \in \calP$ in this response. Note that 
\begin{align}
\sum_{P \in \calP} (x_P - \Delta_P) \geq \sum_{P \in \calP} \barc'_P (x_P - \Delta_P) \geq \sum_{P \in \calP} \barc'_P x_P - B',\label{eq:profit-lb}
\end{align}
where the first inequality follows from the fact that $\barc'_P \leq 1$ for all $P \in \calP$, and the second inequality is a consequence of $\barc'_P\leq\barc_P/c_f$ and $\sum_{P \in \calP} \barc_P \Delta_P / c_{\eprime} \leq B_I /c_{\eprime} = B'$.

Suppose now that $\eprime$ is indeed the most expensive arc that the interdictor touches in his greedy strategy when attacking flow $x$. We also assume that the interdictor's budget does not exceed the cost of stealing all flow; otherwise, from the flow player's perspective, $x$ is a meaningless solution. Observe that in this case $\barc'_P = 1$ for all $P$ with $x_P - \Delta_P > 0$, implying that the first inequality in \eqref{eq:profit-lb} is fulfilled with equality. Note further that $\barc'_P = \barc_P/c_{\eprime}$ for all $P$ with $\Delta_P > 0$, implying that the second inequality is also fulfilled with equality, and therefore the flow player's profit equals the LP value in this case.
\end{proof}

Although $\flowlp$ has a possibly exponential number of variables, the following lemma shows that we can actually solve it in polynomial time.

\smallskip
\begin{lemma}\label{lem:dual-separation}
The linear program $\flowlp$ can be solved in polynomial time.
\end{lemma}
\smallskip

\begin{proof}
Notice that the second term~$B'$ in the objective function is constant and can be ignored for the purpose of the lemma. Consider the dual linear program:
\begin{align*}
\min\quad&\sum_{e\in A}u_e y_e\\
\st\quad&\sum_{e\in P}y_e\geq\barc'_P && \text{for all~$P\in\calP$,}\\
&y_e\geq0 && \text{for all~$e\in A$.}	
\end{align*}
The separation problem can be solved by a series of shortest path computations as follows. For every possible value $\gamma$ of $\barc'_P$ (\ie, for every $\gamma \in \Gamma := \{c'_e \mid e\in A\}$), find a shortest path $P(\gamma)\in\calP$ for the arc weights
\begin{align*}
y^{\gamma}_e:=\begin{cases}
y_e & \text{if $c'_e\geq\gamma$,}\\
\infty & \text{otherwise,}	
\end{cases}
\end{align*}
and denote its length by $\pi(\gamma) := \sum_{e \in P(\gamma)} y^{\gamma}_e$.
We will show that if $\pi(\gamma) \geq \gamma$ for every $\gamma \in \Gamma$, then $y$ is a feasible dual solution. Vice versa, if $\pi(\gamma) < \gamma$ for some $\gamma \in \Gamma$, then the dual constraint associated with $P(\gamma) \in \calP$ is violated.

Suppose in fact that there is a path $P\in\calP$ whose dual constraint is violated, and let $\gamma = \barc'_P$. Observe that this implies $y^{\gamma}_e = y_e$ for all $e \in P$ and thus $\pi(\gamma) < \infty$. Therefore, the shortest path computation with respect to weights $y^{\gamma}$ will find a path $P(\gamma) \in \calP$ containing only arcs with $c'_e \geq \gamma$ and thus $\pi(\gamma) = y^{\gamma}(P(\gamma)) \leq y^{\gamma}(P) = y(P) < \barc'_P = \gamma$.  Conversely, suppose that the length $\pi(\gamma)$ of a shortest path $P(\gamma)$ with respect to the arc weights determined by some $\gamma\in\Gamma$ is such that \mbox{$\pi(\gamma) < \gamma$}. Since $\pi(\gamma)$ is finite, it follows that $c'_e\geq\gamma$ and $y^{\gamma}_e = y_e$ for all $e \in P(\gamma)$. Therefore, $y(P(\gamma)) =  y^{\gamma}(P(\gamma)) = \pi(\gamma) < \gamma\leq \barc'_{P(\gamma)}$, \ie, the dual constraint associated with $P(\gamma)$ is violated.
\end{proof}

\smallskip

\subsubsection*{Solving the LP using Discrete Newton} It is possible to treat $\flowlp$ as a parametric LP, where the objective coefficients $\barc'_p$ depend on the parameter $\lam = 1/c_{\eprime}$.  In these terms, the objective function can be written as $\sum_{P\in\calP} \min\{\lam \barc(P), 1\} x_P - \lam B_I$.  Define $\cmin := \min_{e\in A} c_e$, $\cbot := \max_{P\in\calP} \barc(P)$, and $C := \max_{e\in A} c_e$.  Note that a bottleneck shortest path computation can compute $\cbot$ in $O(m\log n)$ time.  When $\lam\ge 1/\cmin$, then the objective function is $\sum_{P\in\calP} x_P$, \ie, the standard maximum flow problem, and when 
$\lam\le 1/\cbot$, then the objective function is $\lam \sum_{P\in\calP} \barc(P) x_P$.  Thus the region of interest is $\lam\in [1/\cbot, 1/\cmin]$.
This proves that the optimal objective value of $\flowlp$ is a piecewise linear concave function with breakpoints at the inverse of $c_e$ for each value of $c_e\in [\cmin, \cbot]$.  Thus the maximum occurs at one such breakpoint.  Given that it is expensive (though polynomial) to solve the LP, we would like to minimize the number of values of $\lam$ for which we have to solve it.  One way to do this is to use the Newton-$B$ algorithm of \cite{McCormickOrioloPeis2014}, which uses a result of Radzik \cite{Radzik1998} to get a bound of $O(\frac{\log(mC)}{1+\log\log(mC) -
    \log\log(m)})$ LP solves, which is typically much faster than $O(m)$.
It is also possible that the flow player is interested to know what his optimal solution is at all possible breakpoints.  For each $c_e\in [\cmin, \cbot]$ it is easy to compute the range of $B_I$ such that the optimal solution of $\flowlp$ occurs at $c_e$.

\subsubsection*{Relation to multicommodity flow}
While Theorem~\ref{theorem:optimality} and Lemma~\ref{lem:dual-separation} yield an efficient algorithm for computing an optimal strategy for the flow player, this algorithm is based on linear programming.
Aneja et al.~\cite{AnejaChandrasekaranNair2001} also deal with a parametric LP for solving their robust flow model. Unlike in our case however, their LP can be formulated in terms of an \emph{arc flow} and can be solved combinatorially in strongly polynomial time by at most $|V|$ maximum flow computations. Moreover, they also obtain the insight that there is an optimal flow that, before the interdiction, is a maximum flow. Finally, even if the optimal flow needs not to be integer, building upon the concavity of their parametric LP, an optimum integral solution can also be obtained by solving two more maximum flow problems, corresponding to the rounding (up and down) of the fractional optimal value of the parameter.

Motivated by these results, we now discuss whether it is also possible to obtain a combinatorial (strongly polynomial) algorithm for {\flowproblem}.
A first observation is that the path weights $\barc \in \R_+^{\calP}$ do not fulfill the  supermodularity condition (\ref{eq:submod}) for weighted abstract flows. It is also not hard to construct instances where no optimal robust flow $x \in X$ is a maximum flow in the network.
Finally, we establish a close connection of {\flowproblem} to the following feasibility version of the multicommodity flow problem.

\medskip
\begin{description}
\item \textbf{(MF)} Given a directed graph \mbox{$D = (V, A)$}, capacities $u\in \R_+^A$, a finite set of commodities $K$ with demands $d_i$, sources $s_i$, and sinks $t_i$ for $i \in K$, decide whether there is a feasible multicommodity flow satisfying all demands.
\end{description}
\medskip

\noindent
Given an instance $I$ of MF, we construct the following instance $I'$ of {\flowproblem} as follows: We add a super source $s$ with arcs $a_i = (s, s_i)$ for every $i \in K$ and a super sink $t$ with arcs $z_i = (t_i, t)$ for all $i \in K$. For these arcs, we set $u_{a_i} = u_{z_i} = d_i$ and $c_{a_i} = c_{z_i} = i$ for all $i \in K$ (assuming w.l.o.g.~$K = \{1, \dots, k\}$). We further add an additional arc $e^* = (s, t)$ with $u_{e^*} = 1$ and $c_{e^*} = k + 1$. Finally, we set $c_e = \infty$ for all arcs $e \in A$ of the original network and let $B_I = \sum_{i \in K} d_i \cdot i$. 

Note that any $s$-$t$-path $P$ starting with arc $a_i$ and ending with arc $z_j$ fulfills $\barc_{P} = \min \{i, j\}$. This can be used to show that any robust flow of value $1$ induces a feasible multicommodity flow and vice versa, as formalized in the proof of the following lemma.

\smallskip
\begin{lemma}\label{lem:multicommodity-hardness}
There is a feasible solution to instance $I$ of MF if and only if the flow player can achieve a profit of $1$ in the instance $I'$ of {\flowproblem}.
\end{lemma}
\smallskip

\begin{proof}
Assume there is a multicommodity flow in $I$ fulfilling all demands. Consider a decomposition of this flow into a path flow along $s_i$-$t_i$-paths for $i \in K$. Extend each $s_i$-$t_i$-path of this decomposition by the corresponding arcs $a_i$ and $z_i$, obtaining a flow $x$ on $s$-$t$-paths. Further set $x_{P^*} = 1$ for the single-arc path $P^*$ along $e^*$. Let $\calP_i$ denote the subset of $s$-$t$-paths starting with $a_i$ and ending with $z_i$. Observe that $\sum_{P \in \calP_i} x_P = d_i$ and $\barc_P = i$ for all $P\in\calP_i$, $i \in K$. Therefore the interdictor can steal all flow from $x$ except for the $1$ unit along $P^*$.
  
Now assume there is a path flow $x$ for $I'$ with profit $1$ for the flow player. For~\mbox{$i \in K$}, let $\calP'_i$ denote the set of $s$-$t$-paths starting with $a_i$. Observe that $\sum_{P \in \calP'_i} x_P \leq d_i$ and $\barc_P \leq i$ for $P \in \calP'_i$. Therefore
\[
\sum_{i \in K} \sum_{P \in \calP'_i} \barc_P x_P \leq \sum_{i \in K} d_i \cdot i = B_I,
\]
implying the interdictor's budget is always sufficient to steal all flow along all $s$-$t$-paths in $\calP \setminus \{P^*\}$. The flow player's profit can only reach $1$ if the flow on $P^*$ remains untouched, implying that the above inequality must hold with equality. This is only possible if $\sum_{P \in \calP'_i} x_P = d_i$ and $\barc_P = i$ for all $P \in \calP'_i$ with $x_P > 0$. Thus $x_P > 0$ for some $P \in \calP'_i$ implies that $P$ starts with $a_i$ and ends with $z_i$. Therefore, $x$ corresponds to a multicommodity flow in $D$ fulfilling all demands.
\end{proof}
\smallskip

This reduction does not only imply that, in general, optimal solutions to~{\flowproblem} are not integral, but also that a combinatorial strongly polynomial algorithm for {\flowproblem} would lead to a combinatorial strongly polynomial algorithm for MF, the existence of which is a longstanding open problem in combinatorial optimization. As the integral version of MF contains the arc-disjoint path problem as a special case, Lemma~\ref{lem:multicommodity-hardness} also implies hardness of {\flowproblem}'s integral version.

\smallskip
\begin{corollary}
  It is $N\!P$-hard to find an optimal integral solution to {\flowproblem}.
\end{corollary}

\subsubsection*{Generalizations}
The results above can be generalized to a setting where the flow player is also given a budget and needs to find a flow whose cost is within this budget. More precisely, non-negative arc costs~$\gamma \in \R_+^A$ and the flow player budget $B_F$ are additionally given in the input, and the path flow $x$ must fulfill the constraint $\sum_{e \in A} \gamma_e \sum_{P:e \in P} x_P \leq B_F$. The resulting problem still has an efficiently solvable dual separation problem.
In the special case of $\gamma = c$, paying the cost can be interpreted as having to pay for the protection of the flow. 

Further note that Lemma~\ref{lem:dual-separation} and Theorem~\ref{theorem:optimality} are not restricted to single-commodity network flows. In fact, our algorithmic results hold for any set system $\calP$ on some finite ground set, as long as a set $P\in\calP$  minimizing $\sum_{e \in P} y_e$ for given non-negative weights $y \in \R_+^A$ can be computed in polynomial time. This includes the following examples.
\begin{itemize}
\item Let $D$ be a network and let $K$ be a set of commodities, each associated with a source $s_i$ and a sink $t_i$ in $D$. Let $\calP$ be the set of $s_i$-$t_i$-paths in $D$ for all $i \in K$. Note that this multicommodity version of {\flowproblem} can be solved by applying the same separation routine as above to each commodity individually.
\item Let $\calP$ be an abstract network, \ie, a collection of internally ordered subsets fulfilling the switching axiom defined in~\cite{hoffman1974generalization}. In this case, a minimum weight subset $P \in \calP$ can be found using the abstract flow algorithm from~\cite{mccormick1996polynomial} as observed in~\cite{kappmeier2014abstract}.
\item Let $G = (V, E)$ be a graph and let $\calP$ be the set of $b$-matchings in $G$ for some fixed vector \mbox{$b \in \Z_+^V$}. As minimum cost $b$-matchings can be computed in polynomial time, the corresponding version of {\flowproblem}, which is robustly packing $b$-matchings in a capacitated graph, can be solved in polynomial time.
\end{itemize}

\section{A network design variant}\label{sec:design}

We consider the situation where the interdiction costs $c \in \R_+^A$ are not given but have to be determined by the flow player. In the input, we are  given an additional vector $\gamma \in \R_+^A$, such that $\gamma_e c_e$ is the cost that the flow player will pay, for each unit of flow traversing arc $e$, to protect it by an interdiction cost of $c_e$. Furthermore, the flow player is given a budget~$B_F$. 
The flow player again specifies a path flow $x \in X$ and in addition the interdiction costs $c \in \R_+^A$, subject to the budget constraint $\sum_{e \in A} \gamma_e c_{e} \sum_{P \in \calP : e \in P} x_P \leq B_F$ (note that the cost of protecting flow is proportional to its amount).
If $\gamma_e = 0$, we will implicitly assume $c_e = \infty$, prohibiting the interdictor from stealing flow from such arcs. We are thus interested in interdiction costs for the set of vulnerable arcs $A^* := \{e \in A \mid \gamma_e > 0\}$.

The interdictor's budget is still denoted by~$B_I$ and he again chooses a vector $z \in \Omega$ specifying how much flow he steals from each path at each arc, paying a cost of $c_e$ per unit of flow stolen from arc $e \in A$. Note that his optimal strategy is again to proceed greedily starting with a flow-carrying path~$P$ with lowest bottleneck interdiction cost~$\barc_P=\min_{e\in P}c_e$.

We denote the resulting network design problem by {\designproblem}. We will show that this problem can also be solved in polynomial time. The algorithm is based on the insight that there always exists an optimal solution for the flow player with the property that arcs have uniform interdiction costs, in which case the interdictor is indifferent about the possible paths (and arcs) to steal from.

\smallskip
\begin{lemma}\label{lem:uniform-cost}
There exists an optimal solution $(x^*, c^*)$ to {\designproblem} fulfilling
\begin{align*}
c^*_e:=\begin{cases}
\dfrac{B_F}{\Gamma(x^*)} & \text{if there exists $P \in \calP$ with $e\in P$ and $x_P>0$,}\\
0 & \text{otherwise}	
\end{cases}
\end{align*}
for all $e \in A^*$, where $\Gamma(x) := \sum_{P \in \calP} \sum_{e \in P} \gamma_e x_P$.
\end{lemma}
\smallskip

\begin{proof}
Note that we can ignore all arcs $e \in A \setminus A^*$ in context of the lemma. We thus pretend to contract these arcs, replacing every $P \in \calP$ by $P \cap A^*$ and removing all paths $P$ with $P \cap A^* = \emptyset$ from $\calP$ for the remainder of this proof. In particular, this implies $\sum_{e \in P} \gamma_e > 0$ for all $P \in \calP$.

For the sake of the proof, we will consider the relaxation of the problem where the flow player can set different interdiction costs $c_{P, e}$ on the same arc $e$, depending on the path $P$ containing $e$, \ie, the interdictor will have to pay a cost of $c_{P, e}$ for stealing one unit of flow from path $P$ at arc $e$. We will show that for this relaxation there is an optimal solution $(x, c)$ with $c_{P, e} = \frac{B_F}{\Gamma(x^*)}$ if $x_P > 0$, and $c_{P, e} = 0$ otherwise.
Note that this implies $c_{P_1, e} = c_{P_2, e}$ for all $e \in A^*$ and all flow-carrying paths $P_1, P_2 \in \calP$ containing~$e$. Therefore the relaxation yields a feasible solution $(x^*, c^*)$ of the same cost to the original problem {\designproblem}, fulfilling the requirements of the lemma.
  
Concerning the relaxation, we first observe that there always is an optimal solution~$(x, c)$ with $c_{P, e}  = c_{P, \eprime}$ for all $P \in \calP$ and $e, \eprime \in P$, because the interdictor will always strike at the cheapest arc of any path $P$. We therefore denote by $c_P$ the interdiction cost of each arc $e\in P$, and slightly abuse notations by referring to a solution to the relaxation by a pair of vectors $x, c \in \R^{\calP}$ with $x_P$ specifying the flow value and $c_P$ specifying the interdiction cost for path $P \in \calP$.

Let $(x, c)$ be an optimal solution to the relaxation. Without loss of generality, we can assume that $c_P = 0$ implies $x_P = 0$, as the flow player does not gain any profit from paths with $c_P = 0$.
 We will now show that we can change $c$, without decreasing the flow player's profit, in such a way that for every $P \in \calP$ we have $c_P = 0$ if $x_P = 0$, and $c_P = C$ if $x_P > 0$ for some fixed value $C$, yielding therefore a solution to the relaxation that implies the lemma, as discussed above.

Given the interdictor's greedy answer $z \in \Omega$ to $(x, c)$, let $$c' := \max \{c_P \mid \sum_{e \in P} z_{e, P} > 0\}$$ be the cost of the most expensive path touched by the interdictor. Note that $c'$ is also the cost of the most expensive flow-carrying path in~$x$: If there is any path $P$ with $x_P >0$ and $c_P > c'$, since this path is not touched by the interdictor, the flow player might reduce $c_P$ to $c'$ and increase the interdiction cost of all other paths, leading to a higher profit and contradicting the optimality of $(x, c)$. Thus let \mbox{$\calP' := \{P \in \calP \mid c_P = c'\}$} be the set of most expensive paths, and let $\Delta := \sum_{P \in \calP'} \sum_{e \in P} z_{e, P}$ be the total amount of flow stolen by the interdictor from paths in~$\calP'$. Furthermore, define $\gamma_P := \sum_{e \in P} \gamma_e$ and $\Gamma' := \sum_{P \in \calP'} \gamma_P x_P$. Now assume there is another path $P' \in \calP$ with $x_{P'} > 0$ and $0 < c_{P'} < c'$. W.l.o.g., choose $P'$ such that $c_{P'}$ is maximized among all such paths. We investigate the effect of shifting interdiction cost from~$P'$ to the paths in $\calP'$ or vice versa. Decreasing $c_{P'}$ by $\varepsilon$ frees $\gamma_{P'}\varepsilon x_{P'}$ units of the flow player's budget, which can be used to increase uniformly the interdiction cost of all paths in $\calP'$ by $\varepsilon \frac{\gamma_{P'}x_{P'}}{\Gamma'}$. In this case, the interdictor needs $\varepsilon x_{P'}$ budget units less for stealing all flow from $P'$, therefore he can use $c'\Delta + \varepsilon x_{P'}$ budget units for stealing flow from paths in $\calP'$. With the increased interdiction costs for those paths, the amount of flow that remains after the interdictor strikes, \ie, the profit of the flow player, is
  \begin{align}
    f(\varepsilon) := \sum_{P \in \calP'} x_P - \frac{c'\Delta + \varepsilon x_{P'}}{c' + \varepsilon\frac{\gamma_{P'}x_{P'}}{\Gamma'}}.\label{eq:epsilon-change}
  \end{align}
 Note that this is also true for negative values of $\varepsilon$, \ie, increasing $c_{P'}$ by $|\varepsilon|$ and decreasing the interdiction cost of paths in $\calP'$ uniformly by $|\varepsilon| \frac{\gamma_{P'}x_{P'}}{\Gamma'}$. In fact, $f(\varepsilon)$ determines the change in profit for any $\varepsilon$ such that $c_{P'} - \varepsilon \geq 0$ and $c_{P'} + \varepsilon \leq c' - \varepsilon\frac{\gamma_{P'}x_{P'}}{\Gamma'}$, as within this interval all flow on $P'$ will be interdicted before the interdictor touches any path in $\calP'$.
Note that by optimality of $c$ we have
  \[0 = f'(\varepsilon) = - \frac{c'x_P (1 - \frac{\gamma_{P'}}{\Gamma'}\Delta)}{(c' + \varepsilon\frac{\gamma_{P'}x_P}{\Gamma'})^2}\]
  and so $\gamma_{P'}\Delta = \Gamma'$. Inserting this in~\eqref{eq:epsilon-change} yields $f(\varepsilon) = \sum_{P \in \calP'} x_P - \Delta$, \ie, shifting interdiction costs in either direction does not change the flow player's profit. We therefore can choose $\varepsilon$ such that $c_{P'} + \varepsilon = c' - \varepsilon\frac{\gamma_{P'}x_{P'}}{\Gamma'}$, equalizing the interdiction cost of $P'$ and all paths in $\calP'$. This procedure can be repeated until $c_P = C$ for all $P$ with $x_P > 0$ and some constant $C$. It then trivially follows that  $C  = \frac{B_F}{\Gamma(x)}$.
\end{proof}
\smallskip

Recall that we denote the surviving flow after the interdictor applies his greedy strategy $z \in \Omega$ by $\bar{x}$, where $\bar{x}_P = x_P - \sum_{e \in P} z_{e, P}$.
Further observe that given any flow $x \in X$, by choosing $c$ in the same way as defined in the statement of Lemma~\ref{lem:uniform-cost} we obtain a solution $(x, c)$ with profit $\sum_{P \in \calP} \bar{x}_P = \sum_{P \in \calP} x_P -  \frac{\Gamma(x)}{B_F} B_I$. 
It follows that the flow $x^*$ in the optimal solution $(x^*, c^*)$ guaranteed by Lemma~\ref{lem:uniform-cost} must maximize this linear objective.
We obtain the following theorem.

\smallskip
\begin{theorem}
An optimal flow $x \in X$ for the flow player in {\designproblem} can be computed by solving the following linear program:
\begin{align*}
{\uniformcostlp} \qquad \max\quad & \sum_{P \in \calP} \left(1 - \frac{B_I}{B_F} \sum_{e \in P} \gamma_e\right) x_P\\
\st\quad&\sum_{P\in\calP:e\in P}x_P\leq u_e&&\text{for all~$e\in A$,}\\
&x_P\geq0 && \text{for all~$P\in\calP$.}
\end{align*}
\end{theorem}
\smallskip

\begin{corollary}\label{cor:min-cost-flow}
Problem {\designproblem} can be solved in strongly polynomial time. There exists an optimal solution such that $x$ is integral.
\end{corollary}
\smallskip

\begin{proof}
Consider the network $D' = (V, A')$ where $A' := A \cup \{(t, s)\}$ with capacities $u'(e) := u(e)$ for $e \in A$ and $u(t, s) := \infty$. Let $c(e) := \frac{B_I}{B_F}\gamma_e$ for $e \in A$ and $c(t, s) := -1$. It is easy to see that an optimal solution to $\uniformcostlp$ corresponds to a minimum cost circulation in $(D', u', c)$ and vice versa (note that every negative cost cycle in $D'$ contains $(t,s)$). We can thus solve the minimum cost flow problem in $D'$ in strongly~poly\-nomial time and obtain an integral solution to $\uniformcostlp$ and {\designproblem}.
\end{proof}

\smallskip

\subsubsection*{Generalizations and integrality of solutions}
Note that, as in \cref{sec:flow}, the linear program $\uniformcostlp$ can be solved via its dual separation problem in polynomial time for any set system for which a set $P\in\calP$ minimizing $\sum_{e \in P} y_e$ for given non-negative weights $y \in \R_+^A$ can be computed in polynomial time. Therefore, our results for {\designproblem} also hold for all examples of set systems listed at the end of \cref{sec:flow}. However, unlike in the previous section, \cref{cor:min-cost-flow} gives a stronger result for the case that $\calP$ is the set of $s$-$t$-paths in $D$, asserting integrality of the optimal solution and giving a strongly polynomial algorithm. This stronger result also extends to the case where~$\calP$ defines an abstract network, as the path weights $1 - \frac{B_I}{B_F} \sum_{e \in P} \gamma_e$ fulfill~\eqref{eq:submod}, and we can thus solve $\uniformcostlp$ directly by using a combinatorial weighted abstract flow algorithm~\cite{MartensMcCormick2008}.

\section{Increasing existing interdiction costs}\label{sec:combined}

We now consider a generalization of the problem discussed in the previous section, in which initial interdiction costs $c^0$ on the arcs are already present in the input graph, but in which further protection can be obtained by the flow player at the cost of $\gamma_e c^+_e x_e$ for an increase of $c^+_e$ on an arc $e$ with flow~$x_e$. This is motivated by considering network infrastructure which comes with some basic protection for free; in order to get better protection on a particular arc, however, a user has to pay cost proportional to the level of protection and amount of flow he wishes to send. We denote this problem by {\combinedproblem}. The following theorem reveals a surprising jump in complexity as compared to {\designproblem}.

\smallskip
\begin{theorem}\label{thm:hardness}
It is $\NP$-hard to decide whether an instance of {\combinedproblem} has an optimal objective value strictly larger than $0$.
\end{theorem}
\smallskip

\begin{proof}
The proof uses a reduction from the arc-disjoint paths problem: Given a directed graph $D = (V, A)$ and two terminal pairs $s_1, t_1$ and $s_2, t_2$, decide whether there is an $s_1$-$t_1$-path $P_1$ and an $s_2$-$t_2$-path $P_2$ such that $P_1 \cap P_2 = \emptyset$. This problem is NP-hard even in directed acyclic graphs~\cite{even1976complexity}. In the following we assume without loss of generality that~$s_1$ has only one outgoing arc and~$t_1$ has only one incoming arc.

Given an instance of the arc-disjoint path problem, we construct an instance of {\combinedproblem} as follows. We modify the graph by adding a super source~$s$ with arcs $a_1 = (s, s_1)$ and $a_2 = (s, s_2)$ and a super sink $t$ with arcs $z_1 = (t_1, t)$ and $z_2 = (t_2, t)$. Let $M:=|A|+3$ and set 
\begin{align*}
u_{a_1} = u_{z_1} &= 1,& c^0_{a_1} = c^0_{z_1} &= M,& \gamma_{a_1} = \gamma_{z_1} &= \infty,\\ 
u_{a_2} = u_{z_2} &= M,& c^0_{a_2} = c^0_{z_2} &= 1,& \gamma_{a_2} = \gamma_{z_2} &= \infty,\\ 
u_e &= \infty,& c^0_e &= 1,& \gamma_e &= 1, &&\text{for all $e \in A$.}	
\end{align*}
We further set $B_F = |A|(M - 1)$ and $B_I = 2M - 1$.
  
Assume there are arc-disjoint paths $P_1$ from $s_1$ to $t_1$ and $P_2$ from $s_2$ to~$t_2$. The flow player can send $1$ unit of flow along $P'_1 := a_1 \circ P_1 \circ z_1$ and $M$ units of flow along $P'_2 := a_2 \circ P_2 \circ z_2$. Furthermore, he can enforce an interdiction cost of $|A|(M -1)/|P_1| + 1 \geq M$ on all arcs of $P_1$. Note that $\barc_{P'_1} = M$ and $\barc_{P'_2} = 1$ in this solution. The interdictor thus uses $M$ units from his budget to steal all flow from $P'_2$. The remaining $M-1$ units are used to steal $1 - 1/M$ flow units from $P'_1$. As a consequence, the remaining flow after interdiction has value~$1/M>0$ in this case.
  
Assume now that no pair of arc-disjoint paths exists. Let $\calP_2$ denote the set of all $s$-$t$-paths containing $a_2$ or $z_2$ and define $\calP_1 := \calP \setminus \calP_2$. For a given flow player solution with flow $x$ and additional cost $c^+$ define $Y := \sum_{P \in \calP_2} x_P\leq M+1$.  

Let $P'_1 \in \calP_1$, that is, $P'_1$ is of the form~$a_1\circ P_1\circ z_1$ for some~$s_1$-$t_1$-path~$P_1$ in~$D$. As~$P_1$ shares at least one arc with every path in $\calP_2$ (here we use the assumption on the unique outgoing and incoming arc at $s_1$ and $t_1$, respectively), we have
\[(\barc_{P_1} - 1)Y \leq \sum_{e \in P_1} c^+_e \!\!\!\!\!\! \sum_{P \in \calP_2 : e \in P} \!\!\!\!\!\! x_P \leq B_F=|A|(M-1)\enspace.\]
Therefore $\barc_{P_1}\leq|A|(M - 1)/Y + 1$ and thus 
\begin{align}\label{eq:barc_P_1}
\barc_{P'_1} \leq \min \{|A|(M - 1)/Y + 1, \, M\}\qquad \text{for every~$P'_1\in\calP_1$,}
\end{align}
where the upper bound~$M$ stems from the fact that the interdiction cost of $a_1$ and $z_1$ is fixed to $M$ and cannot be increased.
Further note that $\barc_{P'_2} = 1$ for all $P'_2 \in \calP_2$ and the interdictor thus steals all flow from such paths at a total cost of~$Y$.
  If $Y\leq M - 1$, then the interdictor has a budget of at least $M$ left, which is sufficient to steal all flow from paths~$P'_1\in\calP_1$ as the total flow on these paths is at most~$u_{a_1}=1$. 
  
 It remains to consider the case that $M-1<Y\leq M+1$.
Then, due to~\eqref{eq:barc_P_1}, $\barc_{P'_1}< |A| + 1$, for every~$P'_1\in\calP_1$. As the interdictor's remaining budget after stealing all flow from paths in $\calP_2$ is~$2M-1-Y\geq M-2\geq |A|+1$ and the total flow on paths~$P'_1\in\calP_1$ is at most~$1$, also in this case, the interdictor can steal all flow.
  
  Thus any algorithm that is able to distinguish between instances where the optimal value for the flow player is $0$ and instances where the value is positive can also solve the arc-disjoint path problem.
\end{proof}
\smallskip

\begin{corollary}
Unless $P = \NP$, there is no $g(|A|)$-approximation algorithm for {\combinedproblem} for any polynomially computable function $g$.
\end{corollary}

\section{Conclusion}
We presented a new model for robust network flows, in which the interdictor can fractionally remove flow from individual paths. Using structural insights on the interdictor's and flow player's optimal strategies, we obtained polynomial time algorithms for the maximum flow variant as well as for the variant where the flow player additionally determines the protection of his flow. Both these results generalize to a broad class of packing problems.
Somewhat surprisingly, for the variant discussed in the last section, where initial protection infrastructure is already present, we derived a strong inapproximability result. Still, it remains open whether this latter variant of the problem allows for bifactor approximation results in the following sense: Find a solution of value at least $\alpha$ times the optimal surviving flow value, but allow the flow player to exceed his budget $B_F$ by a factor of $\beta_F > 1$, or diminish the interdictor's budget $B_I$ by a factor of $\beta_I < 1$.
Another interesting open question is to investigate in how far our results can be translated to the robust flow model of~\cite{BertsimasNasrabadiStiller2013}, either yielding a better approximation factor or a stronger inapproximability similar to the reduction in the proof of Theorem~\ref{thm:hardness}.

\newpage

\bibliographystyle{siam}
\bibliography{references}

\end{document}